\documentclass[12pt,a4paper]{iopart}
\usepackage{amstext}
\usepackage{graphicx}

\usepackage{iopams}
\usepackage{setstack}
\renewcommand{\vec}[1]{{\boldsymbol{#1}}}

\begin{document}

\title{The extended coupled cluster method and the pairing problem}

\author{Chris Snape and Niels R. Walet }

\ead{Niels.Walet@manchester.ac.uk, Christopher.Snape@manchester.ac.uk}

\address{School of Physics and Astronomy, The University of Manchester, Manchester
M13 9PL, UK}
\begin{abstract}
We study the application of various forms of the coupled cluster method
to systems with paired fermions. The novel element of the analysis
is the study of the breaking and eventual restoration of particle
number in the CCM variants. We specifically include Arponen's extended
coupled cluster method, which describes the normal Hartree-Fock-Bogoliubov
mean field at lowest level of truncation. We show that all methods
converge to the exact results as we increase the order of truncation,
but that the breaking of particle number at an intermediate level
means that this convergence occurs in a surprising way. We argue that
the most straightforward form of the method seems to be the most stable
approach to implement for realistic (large number of particles) pairing
problems
\end{abstract}


\maketitle

\section{Introduction}

The Coupled Cluster Method (CCM) has a long and venerable history
in both Physics and Chemistry. It is a method that can be designed
to incorporate the normal mean-field approximation, but in principle
it also allows one to make systematic improvements, and in principle
converges to the exact solution. After initial work by Coester and
Kümmel \cite{CK60}, and Kümmel and his group \cite{KLZ78} with important
applications in nuclear physics, this was picked up in quantum chemistry
by {\v{C}}i{\v{z}}ek and Paldus \cite{Cizek66,Cizek69,CizekPaldus71,PCS72}
and took off from there (see Ref. \cite{BM07} for a recent review).
About 10 years ago, the method made a renewed appearance in nuclear
physics in the work of Mihaila and Heisenberg \cite{HM99,MH99,MH00,MH00b},
and in a more advanced application in the work of Dean \emph{et al}
\cite{KDH+04,PWG+05,WDG+05,WDG+05b,WGP+05,DHK+05,DGH+05,DHK+06,GPH+06,PDG+06b,HPD+07}.
Atomic nuclei are indeed interesting and complicated, even more so
once we start thinking about heavier open shell nuclei, since we then
have to take into account nuclear superfluidity, described by pairing
of nucleons of opposite spins, as well as the {}``normal'' problems
of nuclear physics (short range repulsion, intermediate attraction,
non-central forces). The normal way to deal with such nuclei using
involves mean-field techniques, but one would like to use a fully
microscopic method such as the CCM to tackle the full many-body problem,
including the pairing part. Since we have some understanding of the
nuclear physics complications, in this paper we shall concentrate
on the pairing problem, to see how the CCM can be applied effectively
to this aspect of the nuclear landscape.

Of course pairing is also extremely important in atomic (fermionic)
condensates, where we have the added advantages that experimentalists
can tune the $s$-wave scattering length, which determines the interaction
in the pairing channel, to almost any value required \cite{Greiner03}.
This has allowed one to study the BEC to BCS transition, and has given
many-body theory access to a whole new set of data that require an
explanation. The extreme elegance and control of the experiments means
that the data requires high accuracy calculations, and is also able
to probe collective modes, some of which are beyond simple mean-field
plus harmonic fluctuations (RPA) calculations. One technique that
gives one easy access to such modes is again the CCM. This gives a
second impetus to this study.

Little work has been done within CCM for problems with pairing forces.
The so-called normal CCM has been used by Emrich and Zabolitzky \cite{EZ84,EZ84b},
but this work seems to have been largely forgotten, and it is probably
not as general as required. If we are interested in systems with strong
correlations, we may want to use methods that go beyond simple mean-field
theory. The coupled cluster method has shown to be a very powerful
technique to make such improvements, but we have little intuition
what the best way to apply the method is. The work of Arponen \cite{ARP83}
and Arponen and Bishop \cite{ABP87b,ABP87,ABP+88,RBA89,AB90,BA90}
provides one way; the use of Brueckner/maximum overlap orbitals \cite{Bru56},
where we start from the BCS wave function, and express all corrections
in terms of BCS quasiparticles, is an alternative.

In this paper we shall investigate both these methods for a simplified
model, with an emphasis on particle number--so we concentrate on a
single one of the aspects that are important for finite nuclei. Before
analyzing this model, let us first describe the methods used succinctly.

\section{The normal CCM}

Let us look at the traditional (normal) Coupled Cluster Method (NCCM)
first, both to make contact with other notations used in the field,
and for later reference. 

We shall use the standard {}``Bochum/Manchester notation'', where
$S$ denotes the cluster operator (called {}``$T$'' in most chemically
inspired literature). We also use the {}``SUB(n)'' notation, where
SUB(1) describes the truncation of the cluster operator to a single
quasiparticle excitation, SUB(2) single plus double (SD), etc. Following
Arponen \cite{ARP83}, we start from a functional,\global\long\def\bra#1{\langle#1|}
\global\long\def\ket#1{|#1\rangle}

\begin{equation}
O(\tilde{s},s)=\bra{\tilde{\phi}}O\ket{\phi}=\bra{\phi_{0}}(1+\tilde{S})e^{-S}Oe^{S}\ket{\phi_{0}},\label{eq:ONCCM}\end{equation}
which is used to evaluate the expectation value of any operator including
the Hamiltonian. This has interesting linking properties: all $S$
operators contract with $O$, and the $\tilde{S}$ has to link with
at least one $S$ or the $O$ operator.

The states used in the expectation value are not Hermitian conjugates,
but are intermediate normalised, \begin{equation}
\bra{\phi_{0}}(1+\tilde{S})e^{-S}e^{S}\ket{\phi_{0}}=1.\end{equation}
The resulting expectation value is a finite polynomial once we impose
the condition\begin{equation}
S=\sum_{I}s_{I}C_{I}^{\dagger},\quad\tilde{S}=\sum_{I}\tilde{s}_{I}C_{I},\label{eq:defC1}\end{equation}
with the generalised annihilation operators \begin{equation}
C_{I}\left|\phi_{0}\right\rangle =0.\label{eq:defC2}\end{equation}
The reason for the polynomial nature of the result is that the nested
commutator expansion \[
e^{-S}Oe^{S}=O+[O,S]+\frac{1}{2!}[[O,S],S]+\ldots\]
terminates at finite order if the operator $O$ contains a finite
number of operators. 

Normal CCM is usually written in a slightly different form; if we
extremise the energy (as expectation value of the Hamiltonian) with
respect to $s_{I}$ and $\tilde{s}_{I}$, we only have to solve for
$s_{I}$:\begin{eqnarray}
\bra{\phi_{0}}C_{I}e^{-S^{\text{eq}}}He^{S^{\text{eq}}}\ket{\phi_{0}} & = & 0,\label{eq:CCMeq}\\
\bra{\phi_{0}}e^{-S^{\text{eq}}}He^{S^{\text{eq}}}\ket{\phi_{0}} & = & E,\label{eq:CCMexp}\end{eqnarray}
where we use {}``eq'' to denote the equilibrium solution of (\ref{eq:CCMeq}).
Of course to calculate the expectation value of any other operator
we require $\tilde{s}_{I}^{\text{eq}}$ as well, which can be found
from a set of inhomogeneous \emph{linear} equations\[
\bra{\phi_{0}}(1+\tilde{S}^{\text{eq}})e^{-S^{\text{eq}}}[H,C_{I}^{\dagger}]e^{S^{\text{eq}}}\ket{\phi_{0}}=0.\]
Finally, by expanding the action\begin{equation}
\mathcal{A}[s,\tilde{s}]=\int dt\bra{\tilde{\phi}}(1+\tilde{S})e^{-S}(H-i\partial_{t})e^{S}\ket{\phi},\label{eq:action_NCCM}\end{equation}
to second order about the equilibrium we get the {}``Harmonic approximation'',
also called EOM-CC \cite{SB84}, with classical action, derived by
shifting $s_{I}\rightarrow s_{I}^{\text{eq}}+s_{I}$\begin{eqnarray}
\mathcal{A} & =\int\biggl\{ i\sum_{I}(\tilde{s}_{I}+\tilde{s}^{\text{eq}})\dot{s}_{I}+\frac{1}{2}s_{I}s_{J}\left[\partial_{s_{I}}\partial_{s_{J}}H(s,\tilde{s}^{\text{eq}})|_{s=s^{\text{eq}}}\right]\nonumber \\
 & +\tilde{s}_{I}s_{J}\left[\partial_{\tilde{s}_{I}}\partial_{s_{J}}H(s,\tilde{s})|_{s=s^{\text{eq}}}\right]\biggr\} dt.\label{eq:A_NCCM_harm-2}\end{eqnarray}
(The linear term, a total derivative, is a boundary term and can thus
normally be removed from the action.) Varying with respect to $\tilde{s}_{I}(t)$,
assuming a harmonic dependence of $s_{I}$, $s_{I}=e_{I}e^{i\omega t}$,
we get\begin{equation}
\left[\partial_{\tilde{s}_{I}}\partial_{s_{J}}H(s,\tilde{s}^{\text{eq}})|_{s=s^{\text{eq}}}\right]e_{J}=\omega e_{I}\quad.\label{eq:harm_NCCM_r}\end{equation}
If we wish to get the tilde eigenstates as well, we need to vary with
respect to $s_{J}(t)$, using $\tilde{s}_{I}=f_{i}e^{i\omega t}$
\begin{equation}
\left[\partial_{s_{I}}\partial_{s_{J}}H(s,\tilde{s}^{\text{eq}})|_{s=s^{\text{eq}}}\right]e_{J}+\left[\partial_{\tilde{s}_{J}}\partial_{s_{I}}H(s,\tilde{s}^{\text{ }})|_{s=s^{\text{eq}}}\right]f_{J}=-\omega f_{I}.\label{eq:harm_NCCM_l}\end{equation}
We can rewrite this as the eigenvectors of what looks like an RPA
excited state problem \cite{RingSchuck} (which is due to the fact
that this is the diagonalisation of a quadratic Hamiltonian) \begin{eqnarray}
\left(\begin{array}{cc}
\partial_{s_{I}}\partial_{s_{J}}H(s,\tilde{s}^{\text{eq}})|_{s=s^{\text{eq}}} & \partial_{\tilde{s}_{J}}\partial_{s_{I}}H(s,\tilde{s}^{\text{ }})|_{s=s^{\text{eq}}}\\
\partial_{s_{J}}\partial_{\tilde{s}_{I}}H(s,\tilde{s})|_{s=s^{\text{eq}}} & 0\end{array}\right)\vec{a} & = & \left(\begin{array}{cc}
0 & -I\\
I & 0\end{array}\right)\omega\vec{a},\end{eqnarray}
where we have used the {}``double height column vector'' \begin{equation}
\vec{a}=\left(\begin{array}{c}
\vec{e}\\
\vec{f}\end{array}\right).\end{equation}
For this particular matrix the eigenvalues are completely determined
by the diagonalisation of the lower left block $\partial_{s_{J}}\partial_{\tilde{s}_{I}}H(s,\tilde{s}^{\text{ }})|_{s=s^{\text{eq}}}$,
which is the matrix of derivatives of the traditional CCM equations,
which are given by the Eqs. (\ref{eq:CCMeq}), and are used to determine
the coefficients $s_{I}$.

\section{The extended CCM}

The extended CCM starts from a similar functional as Eq.~(\ref{eq:ONCCM})
, but now based on a double-exponential form of the bra states, \cite{ARP83}
\begin{equation}
O(\tilde{s},s)=\bra{\tilde{\phi}}O\ket{\phi}=\bra{\phi_{0}}e^{\tilde{S}}e^{-S}Oe^{S}\ket{\phi_{0}}.\label{eq:OECCM}\end{equation}
The operators remain unchanged, i.e., they are defined in Eqs.~(\ref{eq:defC1})
and (\ref{eq:defC2}). Unlike the ordinary NCCM, it has not been widely
applied since it is much more difficult to evaluate. Here we discuss
a quasiparticle approach that seems to hold some promise. We shall
in essence make intimate use of a link to the theory of generalised
coherent states, and the fact that Eq.~(\ref{eq:OECCM}) is identical
to the Dyson-Maleev boson mapping of Lie algebras \cite{KM91} when
we can assign a Lie algebra to the operators $C_{I}$,$C_{I}^{\dagger}$
and their commutators.

When we minimize the energy, we can now no longer separate the steps
in the determination of $s_{I}$ and $\tilde{s}_{I}$, but have to
combine the equations\begin{eqnarray}
\bra{\phi_{0}}C_{I}e^{\tilde{S}^{\text{eq}}}e^{-S^{\text{eq}}}He^{S^{\text{eq}}}\ket{\phi_{0}} & = & 0,\nonumber \\
\bra{\phi_{0}}e^{\tilde{S}^{\text{eq}}}e^{-S^{\text{eq}}}[H,C_{I}^{\dagger}]e^{S^{\text{eq}}}\ket{\phi_{0}} & = & 0,\nonumber \\
\bra{\phi_{0}}e^{\tilde{S}^{\text{eq}}}e^{-S^{\text{eq}}}He^{S^{\text{eq}}}\ket{\phi_{0}} & = & E.\end{eqnarray}
Once we have solved these, we can look for the excited states by solving
the harmonic problem, $s_{I}\rightarrow s_{I}^{\text{eq}}+s_{I}$,
etc.,\begin{eqnarray}
\mathcal{A} & = & \int\biggl\{ i\sum_{I}\bra{\phi_{0}}e^{\tilde{S}}C_{I}^{\dagger}\ket{\phi_{0}}\dot{s}_{I}+\frac{1}{2}s_{I}s_{J}\left[\partial_{s_{I}}\partial_{s_{J}}H(s,\tilde{s}^{\text{eq}})|_{s=s^{\text{eq}}}\right]\nonumber \\
 &  & +\tilde{s}_{I}s_{J}\left[\partial_{\tilde{s}_{I}}\partial_{s_{J}}H(s,\tilde{s}^{\text{eq}})|_{s=s^{\text{eq}}}\right]dt.\biggr\}\end{eqnarray}
Varying with respect to $\tilde{s}_{I}(t)$, assuming a harmonic dependence
of $s_{I}$, $s_{I}=e_{I}e^{i\omega t}$, we get\begin{eqnarray}
\left(\begin{array}{cc}
\partial_{s_{I}}\partial_{s_{J}}H(s,\tilde{s}) & \partial_{\tilde{s}_{J}}\partial_{s_{I}}H(s,\tilde{s}^{\text{ }})\\
\partial_{\tilde{s}_{I}}\partial_{s_{J}}H(s,\tilde{s}) & \partial_{\tilde{s}_{I}}\partial_{\tilde{s}_{J}}H(s,\tilde{s})\end{array}\right)_{\text{eq}}\vec{a}=\mathcal{S}\omega\vec{a},\label{eq:ECCMRPAa}\\
\mathcal{S}=\left(\begin{array}{cc}
0 & -\partial_{\tilde{s}_{J}}\left\langle \phi_{0}\right|e^{\tilde{S}}C_{I}^{\dagger}\left|\phi_{0}\right\rangle \\
\partial_{\tilde{s}_{I}}\left\langle \phi_{0}\right|e^{\tilde{S}}C_{J}^{\dagger}\left|\phi_{0}\right\rangle  & 0\end{array}\right)_{\text{eq}},\label{eq:ECCMRPAb}\end{eqnarray}
where the symplectic matrix $\mathcal{S}$ can be brought to canonical
form by replacing the set of amplitudes $\{\tilde{s}_{J}\}$ by the
new set $\{\sigma_{J}\}$ which solves the nonlinear equations \begin{equation}
\sigma_{I}=\left\langle \phi_{0}\right|e^{\tilde{S}}C_{I}^{\dagger}\left|\phi_{0}\right\rangle ,\end{equation}
and expanding the energy in these new amplitudes. This is not necessary
to solve the RPA, however, and for certain mixed calculations (see
below) such a transformation is not easily done, and we may want to
choose to work with the form given in (\ref{eq:ECCMRPAa},\ref{eq:ECCMRPAb}).

\section{ECCM SUB(1) approximation and HFB}

One of the natural ways to generalise the Hartree-Fock method to problems
with pairing, and the BCS approach to problems with particle-hole
interactions is the Hartree-Fock-Bogoliubov method. We still use a
single Slater-determinant wave function, but the state has no definite
particle number, like the BCS state. As shown below, we can capture
this information in a generalised density matrix \cite{RingSchuck}.
There is a CCM approximation that is equivalent to this state:

The ECCM SUB(1) approximation for a finite fermionic system, where
the reference state is a single Slater determinant of occupied orbitals
$h$ \begin{equation}
\ket{\phi_{0}}=\prod_{h}a_{h}^{\dagger}\ket 0\end{equation}
is defined by the one-body $S$ operator%
\footnote{Thus the SUB($n$) calculation has an $n$-body $S$ operator.%
} \begin{equation}
S=\sum_{ph}s^{ph}a_{p}^{\dagger}a_{h}+\sum_{pp'}s^{pp'}a_{p}^{\dagger}a_{p'}^{\dagger}+\sum_{hh'}s^{hh'}a_{h'}a_{h},\end{equation}
where $h$ labels the single-particle states occupied in the reference
state, and $p$ the unoccupied ones. (In other words, this is the
{}``singles'' truncation of quantum chemistry, adapted to the pairing
problem where we no longer preserve particle number). We now evaluate
(\ref{eq:OECCM}) by first using the nested commutator expansion for
$e^{-S}He^{S}$. In calculating the commutator of $S$ with the Hamiltonian,
either one or both of the operators in $S$ contract with $H$. If
both contract, no further contractions are possible. Having performed
all possible contractions with $S$ we need to now calculate the final
contractions with $\tilde{S}$. In ECCM this will either tie together
two $S$'s, one $S$ and link the other operator to $H$, or link
both its indices to $H$. We thus get a factorisation of $S$ and
$\tilde{S}$ into general objects with two external indices, essentially
one-body densities, which is exactly the Hartree-Fock-Bogoliubov (HFB)
factorisation, if we identify the linked strings of $S$ and $\tilde{S}$'s
with the normal and abnormal densities. This can of course be made
explicit: the algebra is rather lengthy, but see below for a simplified
(BCS) example.

\section{Simplifying SUB(1)}

Let us look at the application of the extended CCM to a pure pairing
problem (without particle-hole excitations) in the SUB1 approximation.
We start from the standard definitions\begin{eqnarray}
S & = & \overbrace{\sum_{p}s_{p}a_{\vec{p}\uparrow}^{\dagger}a_{-\vec{p}\downarrow}^{\dagger}}^{S_{p}}+\overbrace{\sum_{h}s_{h}a_{\vec{h}\downarrow}a_{-\vec{h}\uparrow}}^{S_{h}},\\
\tilde{S} & = & \underbrace{\sum_{p}\tilde{s}_{p}a_{\vec{p}\downarrow}a_{-\vec{p}\uparrow}}_{\tilde{S}_{p}}+\underbrace{\sum_{h}\tilde{s}_{h}a_{\vec{h}\uparrow}^{\dagger}a_{-\vec{h}\downarrow}^{\dagger}}_{\tilde{S}_{h}}.\end{eqnarray}
We also write\begin{eqnarray}
\ket{\phi} & = & e^{S}\ket{\phi_{0}},\qquad\bra{\tilde{\phi}}=\bra{\phi_{0}}e^{\tilde{S}}e^{-S}\quad.\end{eqnarray}
We now analyse the normal and abnormal densities \cite{RingSchuck}
in the standard way.

The normal density $\rho_{i\sigma,j\sigma'}=\bra{\tilde{\phi}}a_{\vec{j}\sigma'}^{\dagger}a_{\vec{i}\sigma}\ket{\phi}$
equals \begin{equation}
\rho_{i\sigma,j\sigma'}=\delta_{\sigma\sigma'}\left[\sum_{\vec{h}}\delta_{\vec{i}\vec{j}}^{h}\left(1-\tilde{s}_{h}s_{h}\right)+\sum_{\vec{p}}\delta_{\vec{i}\vec{j}}^{p}\tilde{s}_{p}s_{p}\right]\quad,\end{equation}
and the abnormal densities require a bit more work, but after some
algebra can be expressed as

\begin{eqnarray}
\kappa_{\vec{i}\sigma,\vec{j}\sigma'} & = & \bra{\tilde{\phi}}a_{\vec{j}\sigma'}a_{\vec{i}\sigma}\ket{\phi}\nonumber \\
 & = & \left[\delta_{\sigma\uparrow}\delta_{\sigma'\downarrow}-\delta_{\sigma\downarrow}\delta_{\sigma'\uparrow}\right]\left\{ \tilde{s}_{h}\delta_{\vec{i},\bar{\vec{j}}}^{h}+s_{p}(1-s_{p}\tilde{s}_{p})\delta_{\vec{i},\bar{\vec{j}}}^{p}\right\} .\end{eqnarray}
The bar denotes the time-reversed state

\subsection{Generalised density}

From the standard block-matrix definition of the generalised density,

\begin{equation}
\mathcal{R}=\left(\begin{array}{cc}
\rho & \kappa\\
-\kappa^{*} & I-\rho^{*}\end{array}\right),\end{equation}
we can easily check that $\mathcal{R}^{2}=\mathcal{R}$. Expanding
out this projector condition, we get the four matrix conditions \begin{eqnarray}
\rho^{2}-\kappa\,\kappa^{*} & = & \rho,\nonumber \\
\rho\kappa-\kappa\rho^{*} & = & 0,\nonumber \\
\rho^{*}\kappa^{*}-\kappa^{*}\rho & = & 0,\nonumber \\
\rho^{*2}-\kappa^{*}\kappa & = & \rho^{*}.\end{eqnarray}
Let us show explicitly how one of these works for the forms found
above. First notice that \begin{eqnarray}
\left[\delta_{\sigma\uparrow}\delta_{\sigma'\downarrow}-\delta_{\sigma\downarrow}\delta_{\sigma'\uparrow}\right]\left[\delta_{\sigma'\uparrow}\delta_{\sigma''\downarrow}-\delta_{\sigma'\downarrow}\delta_{\sigma''\uparrow}\right] & = & -\delta_{\sigma\uparrow}\delta_{\sigma''\uparrow}-\delta_{\sigma\downarrow}\delta_{\sigma''\downarrow}=-\delta_{\sigma\sigma''}\quad.\end{eqnarray}
Thus \begin{eqnarray}
\rho^{2} & = & \delta_{\sigma\sigma'}\left[\sum_{\vec{h}}\delta_{\vec{i}\vec{j}}^{h}\left(1-\tilde{s}_{h}s_{h}\right)^{2}+\sum_{\vec{p}}\delta_{\vec{i}\vec{j}}^{p}(\tilde{s}_{p}s_{p})^{2}\right]\nonumber \\
-\kappa\kappa^{*} & = & \delta_{\sigma\sigma'}\left[\sum_{\vec{h}}\delta_{\vec{i}\vec{j}}^{h}\left(1-\tilde{s}_{h}s_{h}\right)\tilde{s}_{h}s_{h}+\sum_{\vec{p}}\delta_{\vec{i}\vec{j}}^{p}(\tilde{s}_{p}s_{p})\left(1-\tilde{s}_{p}s_{p}\right)\right]\end{eqnarray}
which add up to $\rho$.

Of course, the other three elements can be evaluated in a similar
way. This shows that we have a general non-Hermitian classical mapping
of the general density matrix.

{[}Discuss mapping issues{]}

\subsection{Canonical form}

In the literature (e.g., Ref. \cite{RingSchuck}) we can find the
canonical form for the density matrix as \begin{eqnarray}
\rho & = & v_{p}^{2}\delta_{\vec{p},\vec{p}'}+u_{h}^{2}\delta_{\vec{h},\vec{h}'},\\
\kappa & = & \kappa^{*}=u_{k}v_{k}\delta_{\vec{k},\overline{\vec{k}'}}.\end{eqnarray}
Our expressions should be of the same form, since we have assumed
the diagonal form of $S$. Actually, we pay a price here for unnecessarily
starting with a Hartree-Fock state; all calculations above are valid
with respect to the vacuum as well, where we recover a simpler form
of BCS (which has maximal asymmetry of $\kappa$!). From the relations\begin{eqnarray}
\rho & = & v_{k}^{2}\left(\delta_{\vec{k},\vec{k}'}+\delta_{\overline{\vec{k}},\overline{\vec{k}'}}\right),\\
\kappa & = & \kappa^{*}=u_{k}v_{k}\delta_{\vec{k},\overline{\vec{k}'}},\end{eqnarray}
where we have rewritten the basis as \begin{equation}
\vec{k}=\vec{k}\uparrow,\qquad\overline{\vec{k}}=-\vec{k}\downarrow,\end{equation}
we thus conclude that \begin{eqnarray}
\tilde{s}_{k} & = & v_{k}u_{k}=\frac{1}{2}\sin2\psi_{k},\\
s_{k} & = & v_{k}/u_{k}=\cot\psi_{k}.\end{eqnarray}
The energy, which is equal to the BCS energy, and is thus fully variational,
is

\begin{equation}
I(s,\tilde{s})=2\epsilon_{k}s_{k}\tilde{s}_{k}+v_{k_{1}k_{2}k_{1}k_{2}}s_{k_{1}}\tilde{s}_{k_{1}}s_{k_{2}}\tilde{s}_{k_{2}}+v_{k_{2}k_{2}k_{1}k_{1}}s_{k_{1}}\left(1-s_{k_{1}}\tilde{s}_{k_{1}}\right)\tilde{s}_{k_{2}}.\end{equation}
This needs to be extended with a Lagrange multiplier for particle
number--since we break particle number we need to make sure that at
least the average particle number is correct. At this level if truncation,
we get a term $-\lambda\left(s_{k}\tilde{s}_{k}-N_{0}\right)$ in
the functional.

{[}Needs to be mentioned earlier as well{]}

\subsection{Quasiparticle basis}

It is interesting to note that the parametrisation above is naturally
linked to a bi-canonical operatorial basis ($c^{\dagger}\neq d^{\dagger}$,
but $\{d_{\vec{k}}^{\dagger},c_{\vec{l}}\}=\delta_{\vec{k},\vec{l}}$),
which is obtained by calculating the action of $a_{k}$ on the ket
$\ket{\phi}$ and $a_{k}^{\dagger}$ on the bra $\bra{\tilde{\phi}}$
(the idea to use bi-canonical operators traces back at least as far
as Ref.~\cite{BB69}), in each case subtracting the result from the
initial operator:%
\footnote{i.e., \[
a_{\vec{k}}\left|\phi\right\rangle =s_{k}a_{\overline{\vec{k}}}^{\dagger}\left|\phi\right\rangle \]
and \[
\left\langle \tilde{\phi}\right|a_{\vec{k}}^{\dagger}=\left\langle \tilde{\phi}\right|\left(\tilde{s}_{k}a_{\overline{\vec{k}}}+s_{k}\tilde{s}_{k}a_{\vec{k}}^{\dagger}\right).\]
}\begin{eqnarray}
c_{\vec{k}} & = & a_{\vec{k}}+\left[S,a_{\vec{k}}\right]=a_{\vec{k}}-s_{k}a_{\overline{\vec{k}}}^{\dagger},\nonumber \\
c_{\overline{\vec{k}}} & = & a_{\overline{\vec{k}}}+\left[S,a_{\bar{\vec{k}}}\right]=a_{\overline{\vec{k}}}+s_{k}a_{\vec{k}}^{\dagger},\nonumber \\
d_{\vec{k}}^{\dagger} & = & a_{\vec{k}}^{\dagger}+\left[\tilde{S},a_{\vec{k}}^{\dagger}\right]+\left[S,\left[\tilde{S},a_{\vec{k}}^{\dagger}\right]\right]=(1-\tilde{s}_{k}s_{k})a_{\vec{k}}^{\dagger}-\tilde{s}_{k}a_{\overline{\vec{k}}},\nonumber \\
d_{\overline{\vec{k}}}^{\dagger} & = & a_{\overline{\vec{k}}}^{\dagger}+\left[\tilde{S},a_{\overline{\vec{k}}}^{\dagger}\right]+\left[S,\left[\tilde{S},a_{\overline{\vec{k}}}^{\dagger}\right]\right]=(1-\tilde{s}_{k}s_{k})a_{\overline{\vec{k}}}^{\dagger}+\tilde{s}_{k}a_{\vec{k}}.\end{eqnarray}
This can be inverted to give\begin{eqnarray}
a_{\vec{k}} & = & (1-\tilde{s}_{k}s_{k})c_{\vec{k}}+s_{k}d_{\overline{\vec{k}}}^{\dagger},\nonumber \\
a_{\overline{\vec{k}}} & = & (1-\tilde{s}_{k}s_{k})c_{\overline{\vec{k}}}-s_{k}d_{\vec{k}}^{\dagger},\nonumber \\
a_{\vec{k}}^{\dagger} & = & \tilde{s}_{k}c_{\overline{\vec{k}}}+d_{\vec{k}}^{\dagger},\nonumber \\
a_{\overline{\vec{k}}}^{\dagger} & = & -\tilde{s}_{k}c_{\vec{k}}+d_{\overline{\vec{k}}}^{\dagger}.\end{eqnarray}
Using these relations we can transform the original Hamiltonian (there
is no sum over spin indices, since we have chosen a Hamiltonian with
$S$-wave pairing only):\begin{equation}
H=\sum_{\vec{k},\sigma}\epsilon_{k}n_{\vec{k},\sigma}+\sum_{k_{1}k_{2}k_{3}k_{4}}v_{k_{1}k_{2}k_{3}k_{4}}a_{\vec{k}_{1}}^{\dagger}a_{\overline{\vec{k}}_{2}}^{\dagger}a_{\overline{\vec{k}}_{4}}a_{\vec{k}_{3}}.\end{equation}
If we use a time-reversal invariant potential, $v_{k_{1}k_{2}k_{3}k_{4}}=v_{k_{3}k_{4}k_{1}k_{2}}$,
and we also assume spin-balance, $v_{k_{1}k_{2}k_{3}k_{4}}=v_{k_{2}k_{1}k_{4}k_{3}},$
we have\begin{eqnarray}
H^{\text{qp}} & = & H^{0}+H_{k}^{20}d_{\vec{k}}^{\dagger}d_{\overline{\vec{k}}}^{\dagger}+H_{k}^{20}c_{\overline{\vec{k}}}c_{\vec{k}}+H_{k}^{11}d_{\vec{k}}^{\dagger}c_{\vec{k}}+H_{\bar{k}}^{11}d_{\overline{\vec{k}}}^{\dagger}c_{\overline{\vec{k}}}\nonumber \\
 & + & H_{k_{1}k_{2}k_{3}k_{4}}^{40}d_{\vec{k}_{1}}^{\dagger}d_{\vec{k}_{2}}^{\dagger}d_{\overline{\vec{k}}_{3}}^{\dagger}d_{\overline{\vec{k}}_{4}}^{\dagger}+H_{k_{1}k_{2}k_{3}k_{4}}^{04}c_{\overline{\vec{k}}_{1}}c_{\overline{\vec{k}}_{2}}c_{\vec{k}_{3}}c_{\vec{k}_{4}}\nonumber \\
 & + & H_{k_{1}k_{2}\overline{k}_{3}k_{4}}^{31}d_{\vec{k}_{1}}^{\dagger}d_{\vec{k}_{2}}^{\dagger}d_{\overline{\vec{k}}_{3}}^{\dagger}c_{\vec{k}_{4}}+H_{k_{1}\overline{k}_{2}\overline{k}_{3}\overline{k}_{4}}^{31}d_{\vec{k}_{1}}^{\dagger}d_{\overline{\vec{k}}_{2}}^{\dagger}d_{\overline{\vec{k}}_{3}}^{\dagger}c_{\overline{\vec{k}}_{4}}\nonumber \\
 & + & H_{k_{1}\overline{k}_{2}k_{3}k_{4}}^{13}d_{\vec{k}_{1}}^{\dagger}c_{\overline{\vec{k}}_{2}}c_{\vec{k}_{3}}c_{\vec{k}_{4}}+H_{\overline{k}_{1}\overline{k}_{2}\overline{k}_{3}k_{4}}^{13}d_{\overline{\vec{k}}_{1}}^{\dagger}c_{\overline{\vec{k}}_{2}}c_{\overline{\vec{k}}_{3}}c_{\vec{k}_{4}}\nonumber \\
 & + & H_{k_{1}k_{2}k_{3}k_{4}}^{22}d_{\vec{k}_{1}}^{\dagger}d_{\vec{k}_{2}}^{\dagger}c_{\vec{k}_{3}}c_{\vec{k}_{4}}+H_{\overline{k}_{1}\overline{k}_{2}\overline{k}_{3}\overline{k}_{4}}^{22}d_{\overline{\vec{k}}_{1}}^{\dagger}d_{\overline{\vec{k}}_{2}}^{\dagger}c_{\overline{\vec{k}}_{3}}c_{\overline{\vec{k}}_{4}}\nonumber \\
 &  & \qquad+H_{k_{1}\overline{k}_{2}\overline{k}_{3}k_{4}}^{22}d_{\vec{k}_{1}}^{\dagger}d_{\overline{\vec{k}}_{2}}^{\dagger}c_{\overline{\vec{k}}_{3}}c_{\vec{k}_{4}},\label{eq:Hqp}\end{eqnarray}
which is the non-Hermitian analogue of the standard quasiparticle
Hamiltonian \cite{RingSchuck}. As the normal quasiparticle Hamiltonian,
it can be used to simplify calculations.

\subsection{Higher order terms and Brueckner orbitals}

Of course we can calculate higher order CCM contributions as well;
see below for some explicit examples. In general, one of the main
problems in applying the extended CCM in this way is the proliferation
of terms ({}``diagrams'') containing $S^{(1)}$ contributions--these
describe the rotation of \emph{any} single particle state, external
or internal, whereas in NCCM we only rotate external states. One way
to avoid this proliferation, is to use the quasiparticle Hamiltonian
(\ref{eq:Hqp}) defined above, and express the correlation through
an $S^{(2)}$ operator expressed in the quasiparticles--or for that
fact an even higher-order operator-- \begin{equation}
S^{(2)}=s_{k_{1}k_{2}k_{3}k_{4}}^{(2)}d_{\vec{k}_{1}}^{\dagger}d_{\vec{k}_{2}}^{\dagger}d_{\overline{\vec{k}}_{3}}^{\dagger}d_{\overline{\vec{k}}_{4}}^{\dagger},\qquad\tilde{S}^{(2)}=\tilde{s}_{k_{1}k_{2}k_{3}k_{4}}^{(2)}c_{\overline{\vec{k}}_{1}}c_{\overline{\vec{k}}_{2}}c_{\vec{k}_{3}}c_{\vec{k}_{4}},\end{equation}
i.e.\begin{equation}
I(s^{(2)},\tilde{s}^{(2)},s,\tilde{s})=\left\langle \tilde{\phi}\right|\left\{ \begin{array}{c}
e^{\tilde{S}^{(2)}}\\
1+\tilde{S}^{(2)}\end{array}\right\} e^{-S^{(2)}}H^{\text{qp}}e^{S^{(2)}}\left|\phi\right\rangle .\end{equation}
where the alternative shows that we can have either ECCM or NCCM for
the extra terms. This corresponds to the use of what is called {}``Brueckner
orbitals'' in Quantum Chemistry, as long as we optimize the quasiparticle
operators and the $S^{(n)}$, $n>2$, at the same time. Since the
inclusion of higher order terms is our main target, we shall look
specifically at these two methods.

\section{Single shell pairing}

\subsection{Model investigation}

In order to analyse the various approaches, we study a model. The
simplest model exhibiting many of the features we are interested in,
is the single-shell pairing model. This has a Hamiltonian of the form
\cite{RingSchuck}\begin{equation}
H=-G(\Delta^{\dagger}\Delta-N/2).\end{equation}
In the context used here, we interpret it as describing $\Omega$
states, each with spin up or down%
\footnote{In the original context it has states with angular momentum $j=\Omega-1/2$,
and states of opposite angular momentum projection quantum numbers
form pairs%
}. In both cases we can divide the states into two equal size sets
labeled by $k$ and $\bar{k}$, which are then paired, \begin{eqnarray}
\Delta^{\dagger} & = & \sum_{k=1}^{\Omega}a_{k}^{\dagger}a_{\bar{k}}^{\dagger},\nonumber \\
N & = & \sum_{k=1}^{\Omega}a_{k}^{\dagger}a_{k}+\sum_{k=1}^{\Omega}a_{\bar{k}}^{\dagger}a_{\bar{k}}.\end{eqnarray}

These operators form the well-known SI(2) quasispin algebra \cite{RingSchuck}\begin{eqnarray}
[\Delta,\Delta^{\dagger}] & = & 2(\Omega/2-N/2)\equiv2J_{0},\nonumber \\
{}[J_{0},\Delta^{\dagger}] & = & -\Delta^{\dagger},\label{eq:quasipssin}\\
{}[J_{0},\Delta] & = & \Delta.\nonumber \end{eqnarray}
From the algebra, we can easily derive the result that the non-degenerate
ground state for even particle number $N$ is the state $\Delta^{\dagger N/2}|0\rangle$,
and the energy of this groundstate is \begin{equation}
E_{N}=-G\left(\Omega-\frac{N}{2}\right)\frac{N}{2}.\end{equation}

\subsection{CCM}

In order to find the coupled-cluster approximation to the exact solution,
we need to find the extremum of the constrained energy \begin{equation}
E(s,\tilde{s})=\bra 0e^{-\tilde{S}}e^{S}\left[H-\lambda(N-N_{0})\right]e^{S}\ket 0.\end{equation}
We find to lowest order that $s$ cannot depend on the magnetic quantum
number, since there is no preferred direction, and thus\[
E(s,\tilde{s})=-\lambda\left(2\Omega s\tilde{s}-N_{0}\right)-G\Omega(\Omega-1)s\tilde{s}(1-s\tilde{ns}).\]
Isolating the chemical potential term, we see that \begin{equation}
2\Omega s\tilde{s}=\left\langle N\right\rangle =N_{0},\end{equation}
and thus \begin{equation}
s\tilde{s}=\frac{N_{0}}{2\Omega}.\end{equation}
We see a generic feature of the problem emerging: since we cast a
variational calculation as a bi-variational one, introducing superfluous
variables, there is no unique solution to $s$ and $\tilde{s}$ separately.
The standard choice would be \begin{equation}
\left\langle \Delta\right\rangle =\left\langle \Delta^{\dagger}\right\rangle ,\end{equation}
 where we find\begin{eqnarray}
s & = & \sqrt{\frac{N_{0}}{2\Omega-n}},\nonumber \\
\tilde{s} & = & \sqrt{\frac{N_{0}(2\Omega-N_{0})}{(2\Omega)^{2}}},\end{eqnarray}
but it is not necessary to make this choice. The SUB(1) approximation
to the ground-state energy is thus\begin{equation}
E=-G\frac{N_{0}}{2}\frac{\Omega-1}{\Omega}\left(\Omega-\frac{N_{0}}{2}\right),\end{equation}
 which up to the correction of relative order $1/\Omega$ is the exact
answer.

\subsubsection{{}``Particle'' ECCM}

We shall first study the ECCM based on the normal ({}``particle'')
operators, which, although normally a very lengthy calculation, is
a simple calculation for this model, since we can use the quasispin
algebra (\ref{eq:quasipssin}) to work out all results. In light of
the fact that the exact solutions are of the form $\Delta^{\dagger N/2}|0\rangle$,
we shall only consider the limited set of CCM states with structure\begin{equation}
\ket{\Phi}=\exp\left(\sum_{n=1}^{M}s_{n}\Delta^{\dagger n}\right)\ket 0.\end{equation}
We have performed calculations with $M$ up to 7.

\subsubsection{Quasiparticle ECCM}

The quasiparticle Hamiltonian is quite simple in this case. We use
the quasiparticle number operator $n=d_{k}^{\dagger}c_{k}+d_{\bar{k}}^{\dagger}c_{\bar{k}}$,
and the quasiparticle pair operators $\delta=c_{\bar{k}}c_{k}$ and
$\delta^{\dagger}=d_{k}^{\dagger}d_{\bar{k}}^{\dagger}$ to write
\begin{eqnarray}
H/G= & \Omega & /2-(1-s\tilde{s})s\tilde{s}(\Omega-n)^{2}\nonumber \\
 & + & \left[(1-s\tilde{s})^{2}+(s\tilde{s})^{2}\right]\left[\delta^{\dagger}\delta-(\Omega-n)/2\right]\nonumber \\
 & + & s^{2}\delta^{\dagger}\delta^{\dagger}+\left[\tilde{s}(1-s\tilde{s})\right]^{2}\delta\delta\nonumber \\
 & - & s(1-2s\tilde{s})\delta^{\dagger}(\Omega-n-1)-\tilde{s}(1-s\tilde{s})(1-2s\tilde{s})(\Omega-n-1)\delta.\end{eqnarray}
At the same time\begin{eqnarray}
N & = & -(1-2s\tilde{s})(\Omega-n)+\Omega-2s\delta^{\dagger}-2\tilde{s}(1-s\tilde{s})\delta,\end{eqnarray}
and\begin{eqnarray}
N^{2}= & \Omega^{2} & +4\Omega s\delta^{\dagger}+4\Omega\tilde{s}(1-s\tilde{s})\delta\nonumber \\
 & + & 8(1-s\tilde{s})s\tilde{s}\delta^{\dagger}\delta\nonumber \\
 & + & (1-2s\tilde{s})^{2}(\Omega-n)^{2}+2(2s\tilde{s}(1-s\tilde{s})-(1-2s\tilde{s})\Omega)(\Omega-n)\nonumber \\
 & + & 4s^{2}\delta^{\dagger}\delta^{\dagger}+4\left[\tilde{s}(1-s\tilde{s})\right]^{2}\delta\delta\nonumber \\
 & - & 4s(1-2s\tilde{s})\delta^{\dagger}(\Omega-n-1)-4\tilde{s}(1-s\tilde{s})(1-2s\tilde{s})(\Omega-n-1)\delta.\end{eqnarray}
We now calculate the expectation value of $H-\lambda N$ in the more
general state\begin{equation}
\bra{\tilde{-}}e^{\tilde{S}^{(M)}}e^{-S^{(M)}}\left(H-\lambda N\right)e^{S^{(M)}}\ket -.\end{equation}
where we use the SUB($M$) approximation,\begin{equation}
S^{(M)}=\sum_{n=2}^{M}\frac{1}{n!}s^{(n)}\left(\delta^{\dagger}\right)^{n},\quad\tilde{S}^{(M)}=\sum_{n=2}^{M}\frac{1}{n!}\tilde{s}^{(n)}\delta^{n},\end{equation}
and\begin{equation}
\ket -=e^{S^{(1)}}\left|0\right\rangle ,\quad\bra{\tilde{-}}=\bra{\tilde{0}}e^{\tilde{S}^{(1)}}e^{-S^{(1)}}.\end{equation}
All relevant commutators can easily be done using the quasiparticle
quasispin algebra,

\begin{eqnarray*}
[\delta,(\delta^{\dagger})^{m}] & = & (\delta^{\dagger})^{m-1}(\Omega-n-m-1),\\
{}[n,(\delta^{\dagger})^{m}] & = & 2m(\delta^{\dagger})^{m},\\
{}[\delta n,(\delta^{\dagger})^{m}] & = & m(\delta^{\dagger})^{m-1}(\Omega-n-m+1)(n+2m)+2m(\delta^{\dagger})^{m}\delta,\\
{}[\delta^{2},(\delta^{\dagger})^{m}] & = & m(m-1)(\delta^{\dagger})^{m-2}(\Omega-n-m+2)(\Omega-n-m+1)\\
 &  & +2m(\delta^{\dagger})^{m-1}\delta(\Omega-n-m+2).\end{eqnarray*}
Finally we need\begin{equation}
\bra{\tilde{-}}e^{\tilde{S}}(\delta^{\dagger})^{l}\ket -=\prod_{n=2}\delta_{\sum_{m=2}mk_{m},l}\frac{\left(\tilde{s}^{(n)}\right)^{k_{n}}}{k_{n}!}\frac{l!\Omega!}{(\Omega-l!)}.\end{equation}

\subsection{Results}

\subsubsection{ECCM based on particles}

We first wish to analyse the convergence of the ECCM based on the
original operators. This has two reasons. First of all this will be
a good test of the technology used in the implementation of the equations,
which are all based on the algebraic implementation of the quasispin
algebra--we shall see below why this is important. Also, we would
like to understand what type of solutions we get, and how we approach
the exact solution.

\begin{figure}
\begin{centering}
\includegraphics[clip,width=10cm]{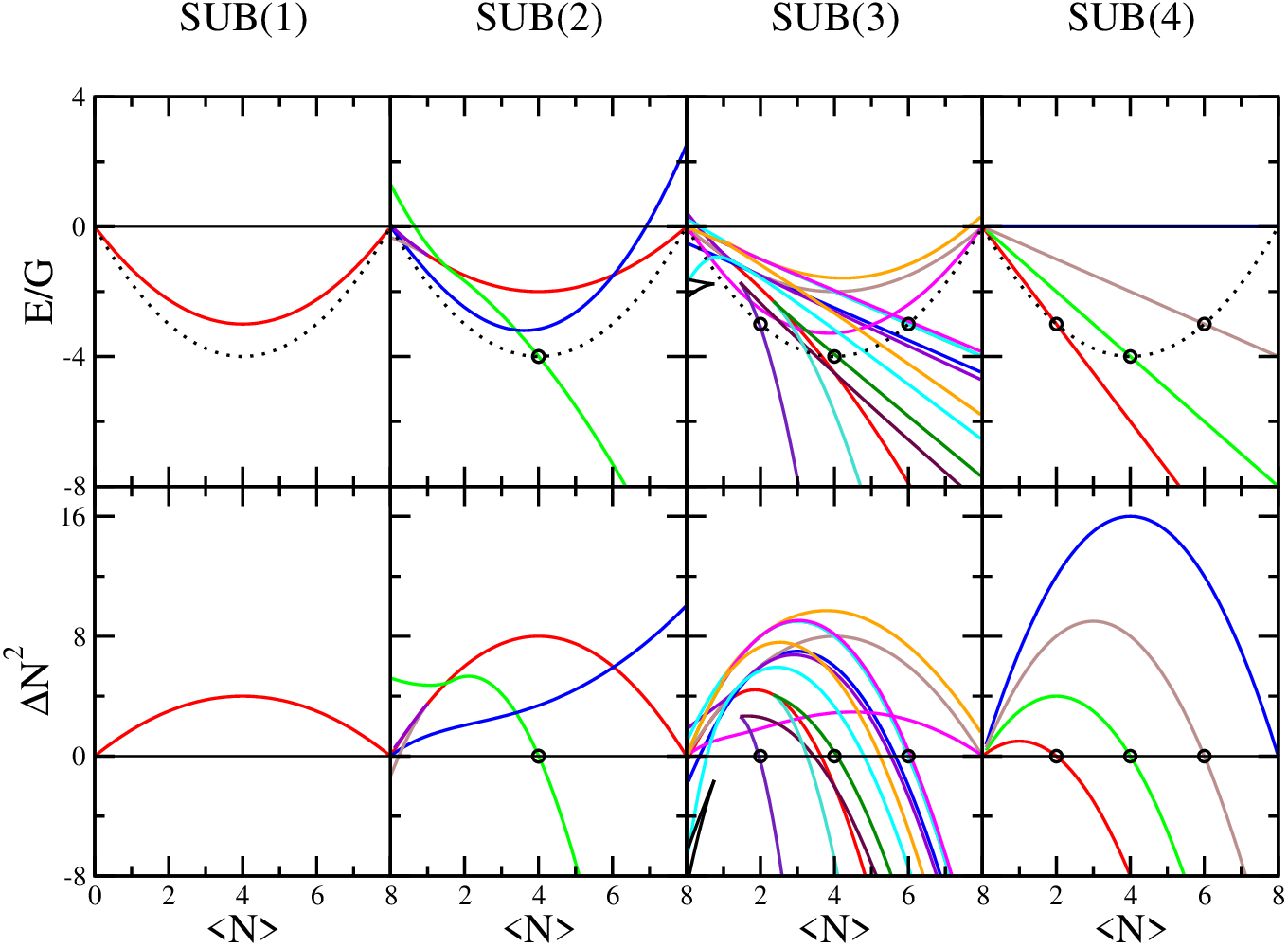}
\par\end{centering}

\caption{\label{fig:all4}A plot of the energy (upper panels) and the fluctuation
in particle number (lower panels) for SUB(1) to SUB(4) ECCM based
on the particle picture, for $\Omega=4$. Black circles denote an
exact solution.}

\end{figure}

With the help of Mathematica, we can obtain the full solution to the
CCM equations, i.e., all \emph{rea}l solutions to the coupled polynomial
equations, see Fig.~\ref{fig:all4}. As we increase the truncation
from the SUB(1) level--which is just the normal BCS approximation--
to the SUB(2) level we see some new solutions appearing. The green
line, which is not a very good solution in most cases, is actually
exact at half filling. The blue line is a better approximation than
BCS up to about half filling. The SUB(3) results give a rather busy
picture, with lots of solutions, among which are exact solutions (for
$N=0,2,4,6,8$). Finally for SUB(4), where ECCM coincides with NCCM,
we find exactly four branches of the solution, but each physical solution
lies on a different branch. Also, all of these solutions are infinitely
degenerate; something which is not obvious from the diagrams! This
should not surprise us: in the SUB(4) calculation each of these states
is orthogonal to the reference state, and is (essentially) generated
by a single $\Delta^{\dagger n}$ excitation on the ground state.

In general we can prove that \begin{equation}
\langle H\rangle=\frac{1}{4}\langle N^{2}\rangle-\frac{\Omega}{2}\langle N\rangle.\end{equation}
Putting in the explicit form of the ground state energy, we can show
that a state where the expectation value of $H$ is the ground-state
energy is also a state where $\langle N^{2}\rangle=\langle N\rangle^{2}$,
i.e., $\Delta N^{2}=0$.

It should worry us that the usual implicit assumption in CCM--that
the physical solution is smoothly connected in parameter space--is
not correct in this case. If we are interested in a relatively low
order SUB($n$) calculation for a problem with many particles this
may not be such an important problem, but this requires detailed investigation. 

\begin{figure}
\begin{centering}
\includegraphics[clip,width=14cm]{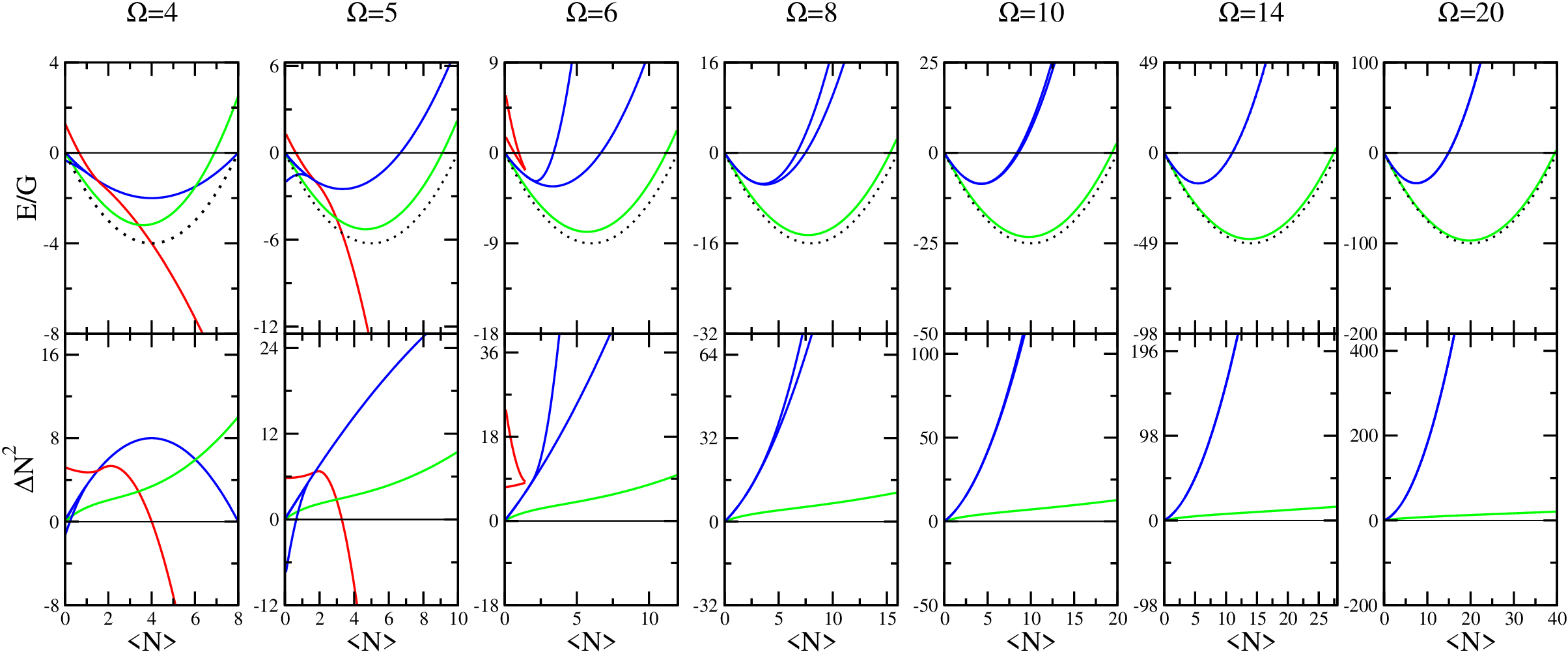}
\par\end{centering}

\caption{\label{fig:part_eccm}A plot of the energy (upper panels) and the
fluctuation in particle number for SUB(2) ECCM based on the particle
picture, for a variety of values of the degeneracy $\Omega$.}

\end{figure}

Notwithstanding those concerns, if we apply the SUB(2) ECCM to a variety
of values of $\Omega$, see Fig.~\ref{fig:part_eccm}, we see that
this is quite a successful and stable improvement to the BCS below
the half-filled shell. Above that we could work with holes relative
to the full shell and get much better results.

\begin{figure}
\begin{centering}
\includegraphics[clip,width=10cm]{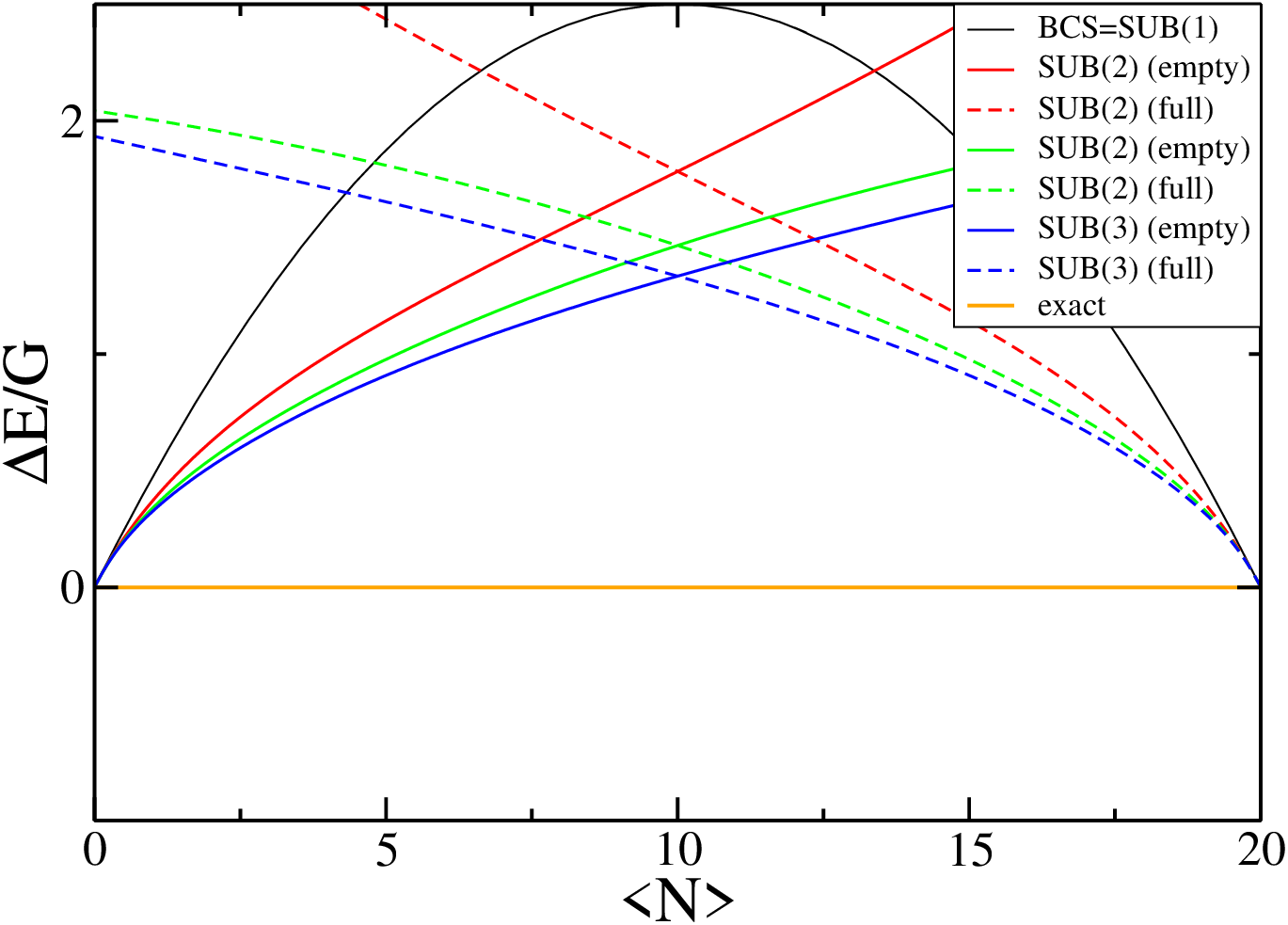}
\par\end{centering}

\caption{\label{fig:Evsnc}A plot $\Delta E=E-E_{\text{exact}}$ vs $\langle N\rangle$
for $\Omega=10$ using ECCM, with {}``ordinary'' operators, rather
than quasiparticle ones. We compare different levels of truncation,
with either hole operators acting on the completely filled state (full)
or particle operators acting on the empty state.}

\end{figure}

We can now concentrate on the {}``physical branch'' of solutions,
identified by the smooth connection to the ground state, and use solution
following techniques to trace the higher order solutions for this
value. In other word, we only solve a small subset. We choose $\Omega=10$,
and find the result of Fig. \ref{fig:Evsnc}. We see that we can indeed
improve the results considerably, but also note the indication that
there is a limit to the improvement. This is obvious from the fact
that we know this is not the exact solution as we increase the truncation
level.

\subsubsection{Brueckner orbitals}

\begin{figure}
\begin{centering}
\includegraphics[clip,width=10cm]{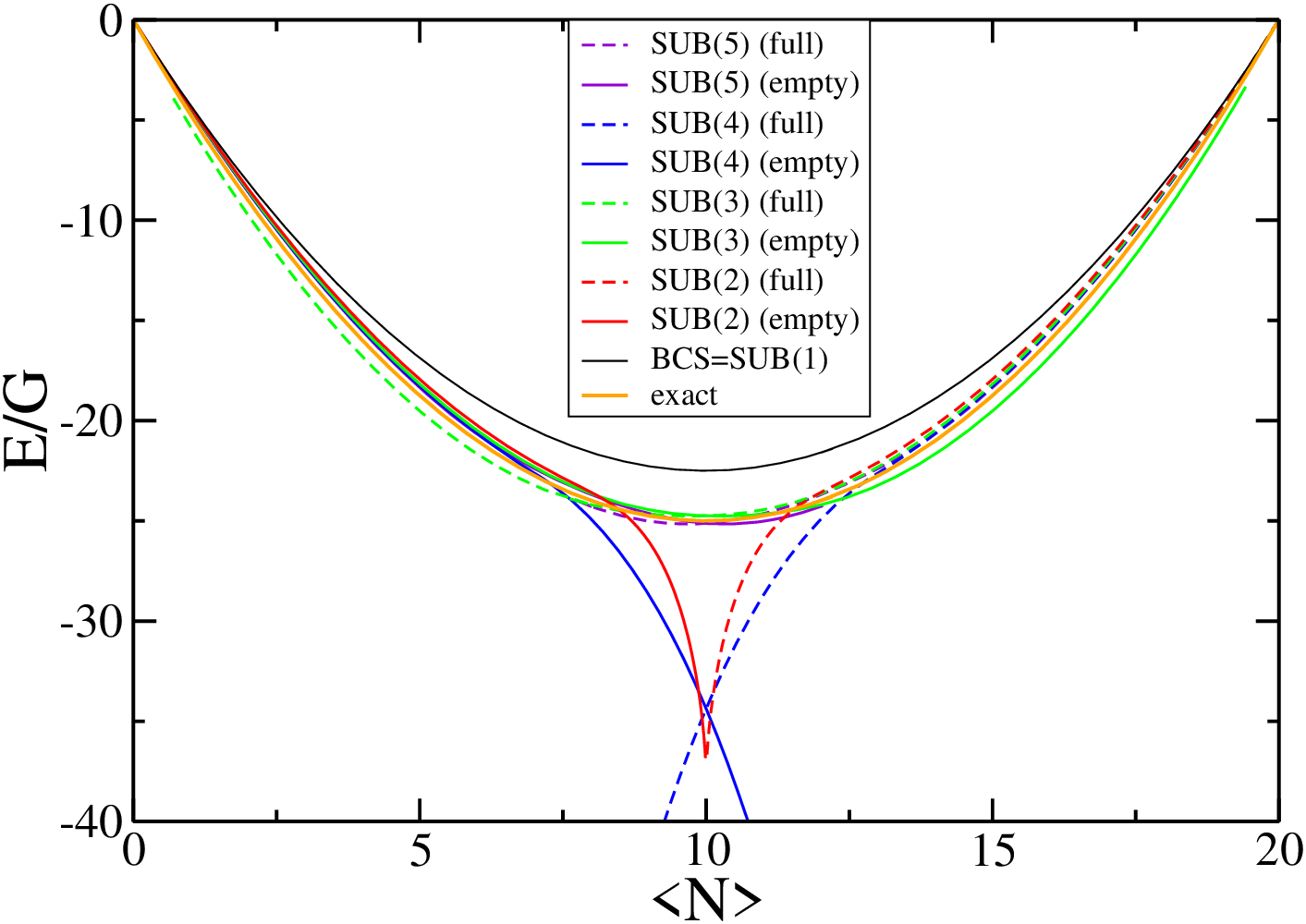}
\par\end{centering}

\caption{\label{fig:Evsn}A plot $E$ as a function of $\langle N\rangle$
for $\Omega=10$, using the quasispin ECCM. We compare different levels
of truncation, with either the filled (full) or empty state as a reference
state.}

\end{figure}

We would like to analyse the same systems for the quasiparticle ECCM
(and NCCM). Unfortunately getting a full set of solutions to the quasiparticle
ECCM, even for $\Omega=4$, as in Fig. \ref{fig:all4} has proven
to be impossible-the equations are just too complicated to get a complete
set of solutions beyond SUB(2). If we look at the quasiparticle NCCM
and ECCM for the SUB(2) case (only $s^{(2)}$ and $\tilde{s}^{(2)}$are
non-zero) we get a rather interesting looking set of solutions, see
Fig.~\ref{fig:Evsn}.

\begin{figure}
\begin{centering}
\includegraphics[clip,width=14cm]{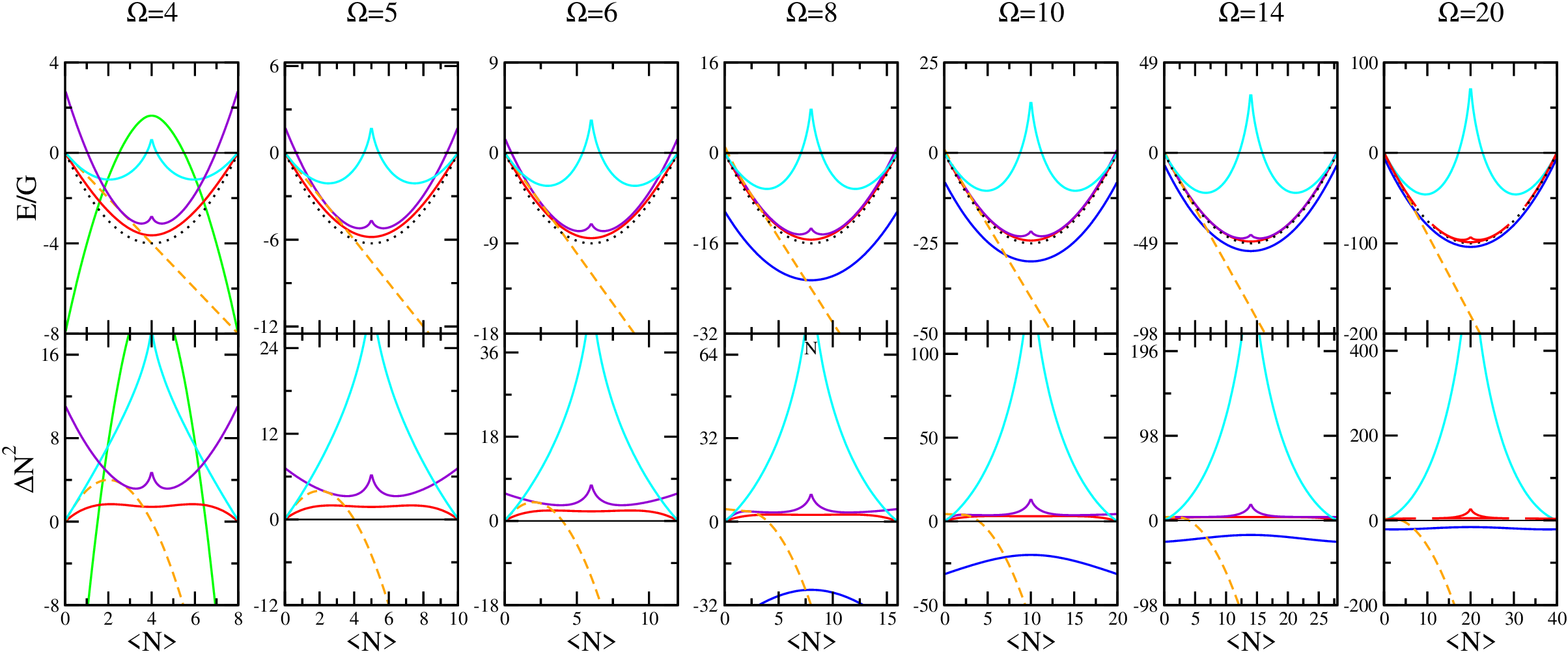}
\par\end{centering}

\caption{\label{fig:qp_nccm}A plot of the energy (upper panels) and the fluctuation
in particle number (lower panel) for SUB(2) quasiparticle NCCM based
on the quasiparticle picture, for a variety of values of the degeneracy
$\Omega$.}

\end{figure}

\begin{figure}
\begin{centering}
\includegraphics[clip,width=14cm]{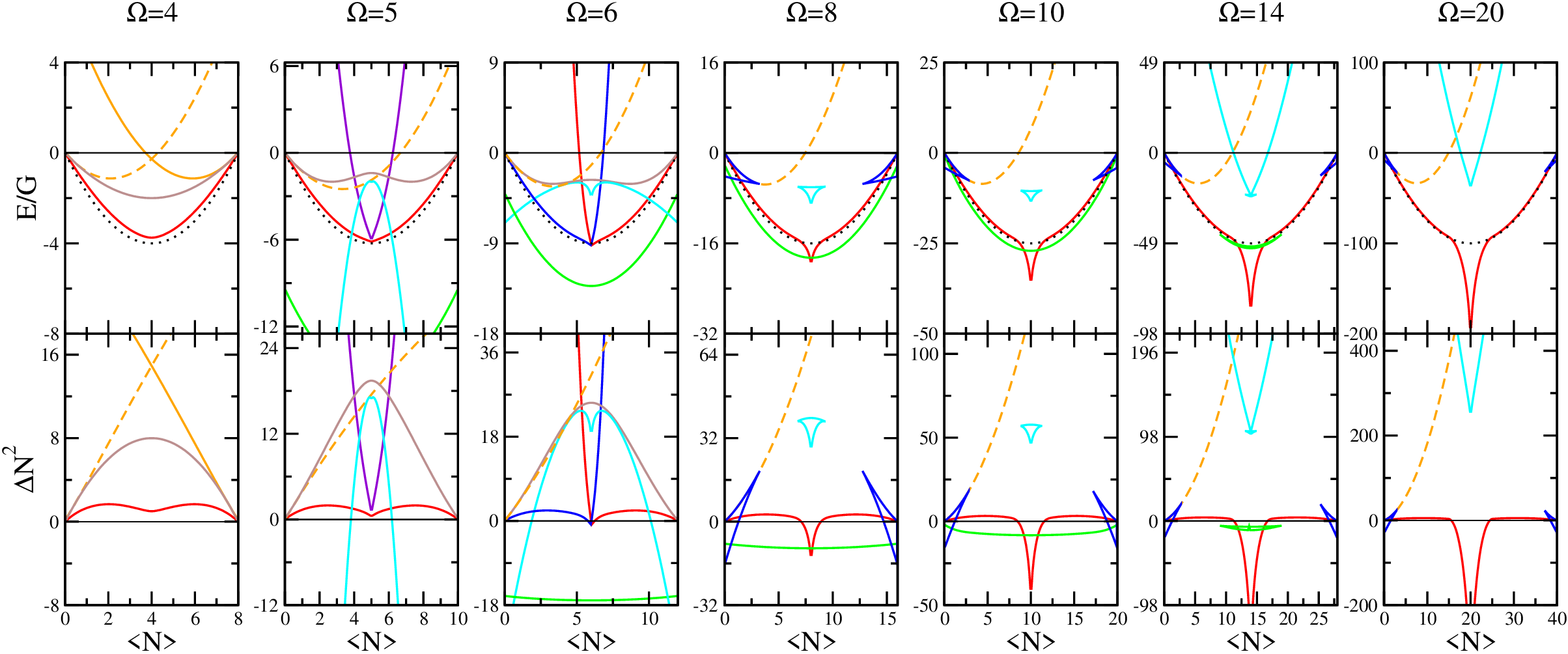}
\par\end{centering}

\caption{\label{fig:qp_eccm}A plot of the energy (upper panels) and the fluctuation
in particle number for SUB(2) quasiparticle ECCM  based on the quasiparticle
picture, for a variety of values of the degeneracy $\Omega$.}

\end{figure}

In that figure we see the usual multitude of solutions; the physically
correct one (that turns to $E=0$ for $\langle N\rangle=0$) collapses
at mid-shell. This agrees with the fact that $\Delta N^{2}$ goes
negative around midshell, showing this is or has become an unphysical
solution. Doing the same calculation with NCCM for the $S^{(2)}$
operators gives a better result, to our great surprise.

If we increase the order of the calculation, we see a suggestion of
convergence for the odd (but not even orders).%
\begin{figure}
\begin{centering}
\includegraphics[clip,width=10cm]{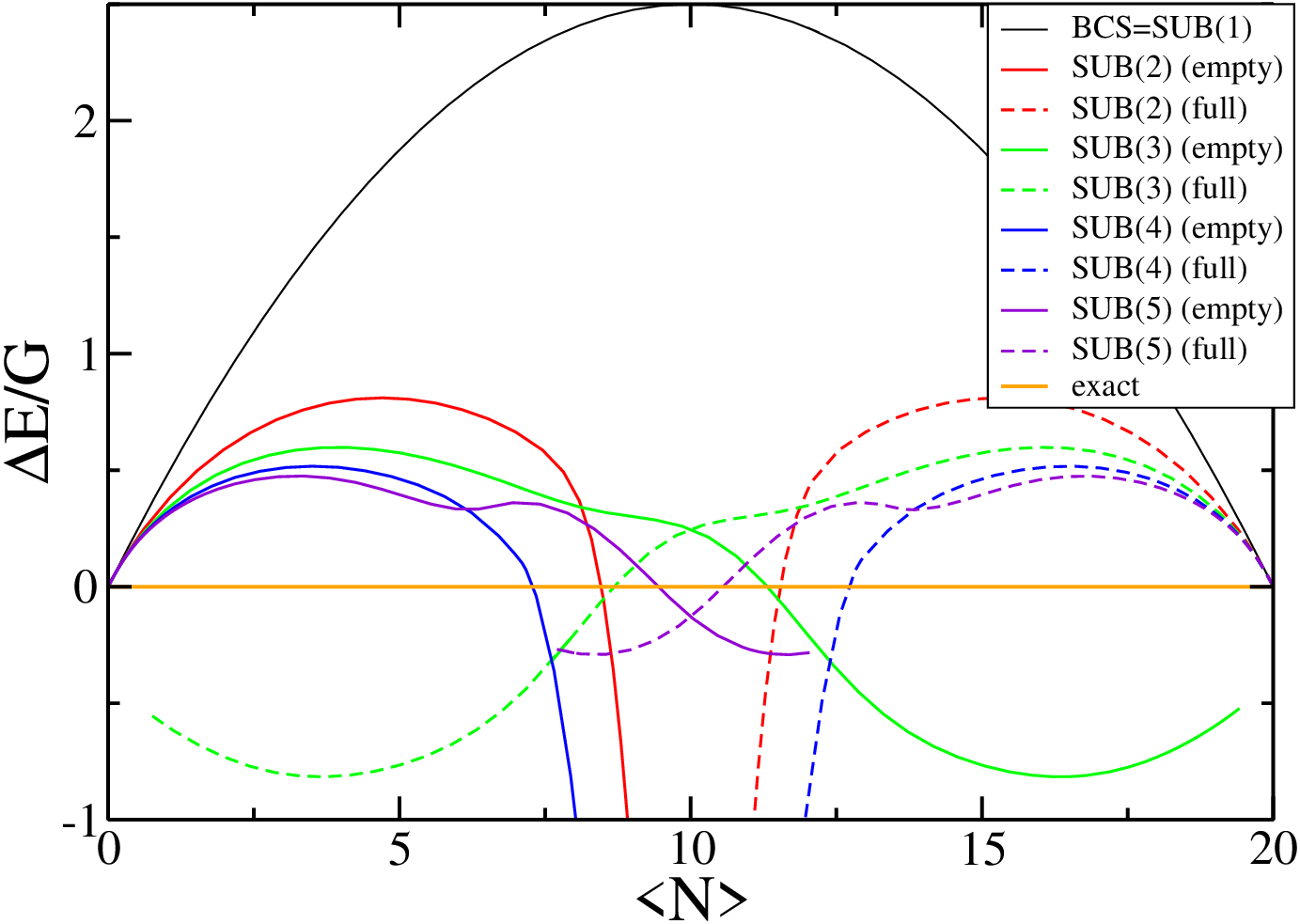}
\par\end{centering}

\caption{\label{fig:Evsnb}A plot $\Delta E=E-E_{\text{exact}}$ vs $\langle N\rangle$
for $\Omega=10$ , using the quasispin ECCM. We compare different
levels of truncation, with either the filled (full) or empty state
as a reference state.}

\end{figure}

\begin{figure}
\begin{centering}
\includegraphics[clip,width=10cm]{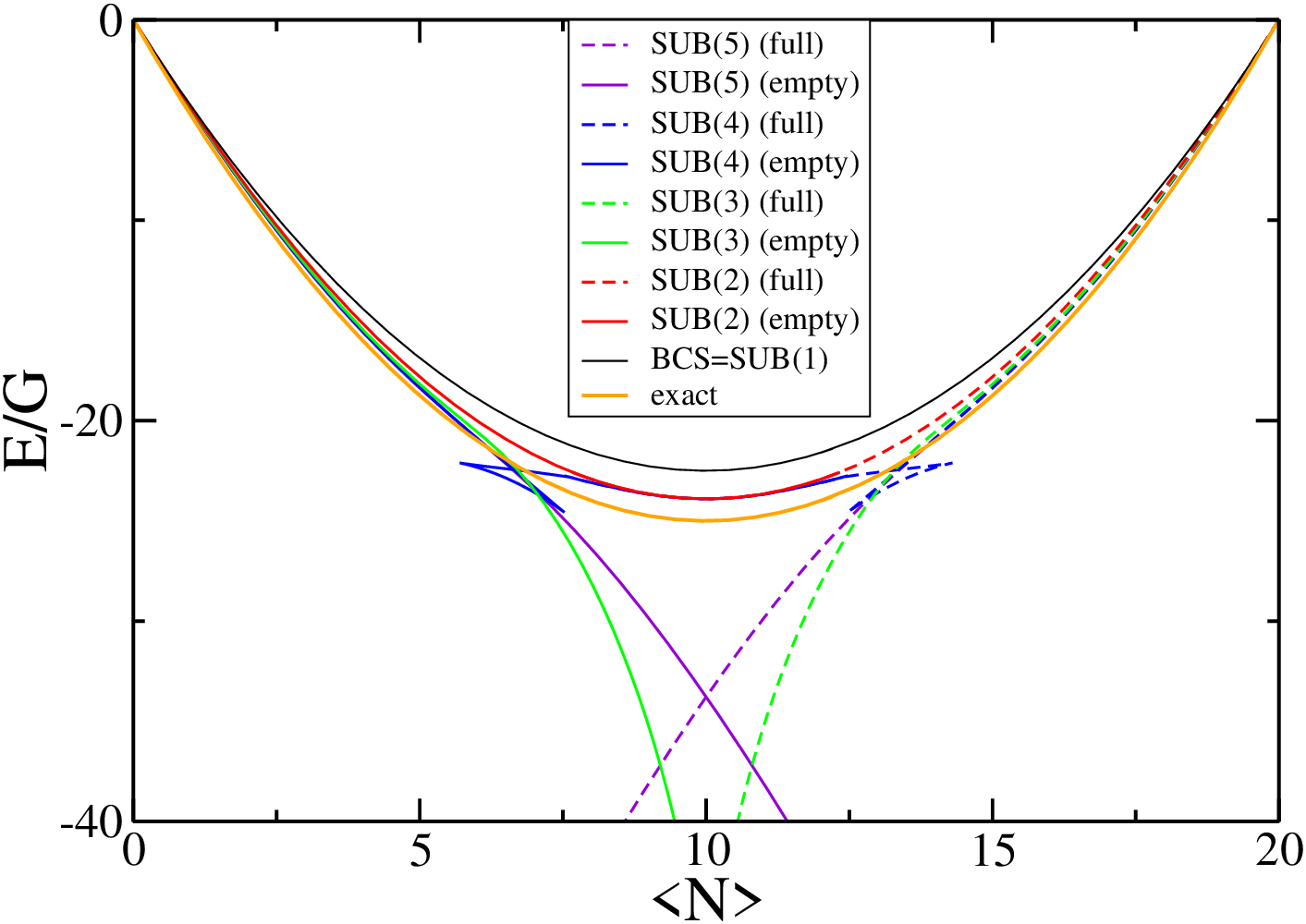}
\par\end{centering}

\caption{\label{fig:Evsn_nccm}A plot $E$ vs $n$ for $\Omega=10$, using
the quasispin NCCM. We compare different levels of truncation, with
either the filled (full) or empty state (empty) as a reference state.}

\end{figure}
We can perform a similar NCCM calculation. Here we can show that the
SUB(2) calculation in the restricted space is exact (the full solution
collapses to the {}``two-pair'' form assumed in all the above calculations).
The solution collapses in higher order, with the strangest result
at order 4. The sharp turns are real--our solution following technique
resolves these exactly.

\subsubsection{Maximum overlap orbitals}

We can finally try to provide one more set of solutions, where we
fix the SUB(1) states by a fixed mean-field calculation, and only
allow SUB(2) excitations beyond that (if we allow further SUB(1) excitations,
we are back to the case of Brueckner orbitals). Unfortunately, there
are no solutions to the resulting equations, i.e., we can't both have
the particle number correct at mean-field and at quasiparticle SUB(2)
level.

\subsubsection{Comparison of methods}

All of this raises a couple of questions:
\begin{enumerate}
\item Should we use (trust) the ECCM or the NCCM based calculations--we
have no clear answer at the moment
\item Should we use the quasispin approach, or should we stick to the ordinary
ECCM without quasiparticles? This can be answered, since the quasispin
algebra make it easy to a calculation with \[
e^{\sum_{n}S^{(n)}\Delta^{\dagger n}}\ket 0\]
and similar for the full shell. The results, shown in Fig.~\ref{fig:Evsnc},
suggest that there is no gain in using this approach, but in general
the approach using the quasiparticles is much more compact--especially
if we stick to the NCCM truncation.
\end{enumerate}
Before drawing too strong a conclusion, we need to look at the harmonic
fluctuations, which in this case is the solution to the generalised
RPA.

\subsubsection{Harmonic excitations}

In order to look at the stability of the solutions and excitations,
we must calculate the harmonic fluctuations about the equilibrium,
also called the RPA for the Brueckner orbitals (the particle case
is trivial, and was already discussed above). The easiest way to derive
this is to use the time-dependent variational principle, derivable
from the action\begin{equation}
\mathcal{A}[s,\tilde{s}]=\int dt\langle\tilde{\phi}|e^{\tilde{S}}e^{-S}(H-i\partial_{t})e^{S}|\phi\rangle.\label{eq:action}\end{equation}
The only tricky point is that $\phi$ is time-dependent, as are the
\emph{operators} appearing in $S$. The expectation value of $H$
has already been evaluated, and we can easily find that (we use {}``NC''
as a short-hand for the nested commutator expansion)\begin{eqnarray}
\langle\tilde{\phi}|e^{\tilde{S}}e^{-S}(i\partial_{t})e^{S}|\phi\rangle & = & \langle\tilde{\phi}|e^{\tilde{S}}e^{-S}(i\partial_{t}e^{S})|\phi\rangle+\langle\tilde{\phi}|e^{\tilde{S}}i\partial_{t}|\phi\rangle\nonumber \\
 & = & \langle\tilde{\phi}|e^{\tilde{S}}\sum_{m=2}\frac{1}{m!}(i\partial_{t}s^{(m)})\left(\delta^{\dagger}\right)^{m}|\phi\rangle\nonumber \\
 &  & +\langle\tilde{\phi}|e^{\tilde{S}}\sum_{n=1}\frac{1}{n!}\left[\sum_{m=2}\frac{1}{m!}s^{(m)}i\partial_{t}\left(\delta^{\dagger}\right)^{m},S\right]_{n-1}|\phi\rangle\nonumber \\
 &  & \quad+\langle\tilde{\phi}|e^{\tilde{S}}i\dot{s}\Delta^{\dagger}|\phi\rangle.\end{eqnarray}
The first term can be trivially calculated with the techniques we
have developed; the other two require (a bit of) further algebra.
Let us start with the final term. We need to express $\Delta$ in
quasispin operators,\begin{equation}
\Delta^{\dagger}=\sum_{k}a_{k}^{\dagger}a_{\bar{k}}^{\dagger}=\sum_{k}\left(\tilde{s}c_{\bar{k}}+d_{k}^{\dagger}\right)\left(-\tilde{s}c_{k}+d_{\bar{k}}^{\dagger}\right)=-\tilde{s}^{2}\delta+\delta^{\dagger}-\tilde{s}(n-\Omega).\end{equation}
We thus find that \begin{equation}
i\dot{s}\langle\tilde{\phi}|e^{\tilde{S}}\Delta^{\dagger}|\phi\rangle=\Omega\tilde{s}\dot{s},\end{equation}
since $\tilde{S}$ only contains terms with two or more $\delta$'s.

We are left with the middle operator, which requires most work. First
of all\begin{eqnarray}
\partial_{t}\delta^{\dagger} & = & \partial_{t}\sum_{k}\left[\left(1-s\tilde{s}\right)a_{k}^{\dagger}-\tilde{s}a_{\bar{k}}\right]\left[\left(1-s\tilde{s}\right)a_{\bar{k}}^{\dagger}+\tilde{s}a_{k}\right]\nonumber \\
 & = & -(\dot{s}\tilde{s}+\dot{\tilde{s}}s)\left[a_{k}^{\dagger}d_{\bar{k}}^{\dagger}+d_{k}^{\dagger}a_{\bar{k}}^{\dagger}\right]+\dot{\tilde{s}}\left[-a_{\bar{k}}d_{\bar{k}}^{\dagger}+d_{k}^{\dagger}a_{k}\right]\nonumber \\
 & = & -(\dot{s}\tilde{s}+\dot{\tilde{s}}s)\left[\left(\tilde{s}c_{\bar{k}}+d_{k}^{\dagger}\right)d_{\bar{k}}^{\dagger}+d_{k}^{\dagger}\left(-\tilde{s}c_{k}+d_{\bar{k}}^{\dagger}\right)\right]\nonumber \\
 & \quad & +\dot{\tilde{s}}\left[-\left[\left(1-s\tilde{s}\right)c_{\bar{k}}-sd_{k}^{\dagger}\right]d_{\bar{k}}^{\dagger}+d_{k}^{\dagger}\left[\left(1-s\tilde{s}\right)c_{k}+sd_{\bar{k}}^{\dagger}\right]\right]\nonumber \\
 & = & \left[-2(\dot{s}\tilde{s}+\dot{\tilde{s}}s)+2\dot{\tilde{s}}s\right]d_{k}^{\dagger}d_{\bar{k}}^{\dagger}\nonumber \\
 &  & +\left[-(\dot{s}\tilde{s}+\dot{\tilde{s}}s)\tilde{s}-\dot{\tilde{s}}\left(1-s\tilde{s}\right)\right]c_{\bar{k}}d_{\bar{k}}^{\dagger}\nonumber \\
 &  & +\left[(\dot{s}\tilde{s}+\dot{\tilde{s}}s)\tilde{s}+\dot{\tilde{s}}\left(1-s\tilde{s}\right)\right]d_{k}^{\dagger}c_{k}\nonumber \\
 & = & -2\tilde{s}\dot{s}\delta^{\dagger}+\left(\dot{\tilde{s}}+\tilde{s}^{2}\dot{s}\right)(n-\Omega).\end{eqnarray}
This can be used to find that\begin{eqnarray}
\partial_{t}\left(\delta^{\dagger}\right)^{m} & = & \sum_{l=0}^{m-1}\left(\delta^{\dagger}\right)^{l}\left[-2\tilde{s}\dot{s}\delta^{\dagger}+\left(\dot{\tilde{s}}+\tilde{s}^{2}\dot{s}\right)(n-\Omega)\right]\left(\delta^{\dagger}\right)^{m-l-1}\nonumber \\
 & = & -2\tilde{s}\dot{s}m\left(\delta^{\dagger}\right)^{m}\nonumber \\
 &  & \quad+\left(\dot{\tilde{s}}+\tilde{s}^{2}\dot{s}\right)\sum_{l=0}^{m-1}\left(\delta^{\dagger}\right)^{l}\left(\delta^{\dagger}\right)^{m-l-1}\left[n+2(m-l-1)-\Omega\right]\nonumber \\
 & = & -2\tilde{s}\dot{s}m\left(\delta^{\dagger}\right)^{m}+\left(\dot{\tilde{s}}+\tilde{s}^{2}\dot{s}\right)\left(\delta^{\dagger}\right)^{m-1}\sum_{k=0}^{m-1}\left[n-\Omega+2k\right]\nonumber \\
 & = & -2\tilde{s}\dot{s}m\left(\delta^{\dagger}\right)^{m}+\left(\dot{\tilde{s}}+\tilde{s}^{2}\dot{s}\right)\left(\delta^{\dagger}\right)^{m-1}m\left[n-\Omega+m-1\right]\quad.\end{eqnarray}
This can be used to evaluate\begin{eqnarray}
\lefteqn{\langle\tilde{\phi}|e^{\tilde{S}}\sum_{n=1}\frac{1}{n!}\left[\sum_{m=2}\frac{1}{m!}s^{(m)}i\partial_{t}\left(\delta^{\dagger}\right)^{m},S\right]_{n-1}|\phi\rangle}\nonumber \\
 & = & i\sum_{m=2}\frac{1}{m!}s^{(m)}\bigl[\left(-2\tilde{s}\dot{s}m\right)\langle\tilde{\phi}|e^{\tilde{S}}\left(\delta^{\dagger}\right)^{m}|\phi\rangle\nonumber \\
 & \quad & +\left(\dot{\tilde{s}}+\tilde{s}^{2}\dot{s}\right)m(-\Omega+m-1)\langle\tilde{\phi}|e^{\tilde{S}}\left(\delta^{\dagger}\right)^{m-1}|\phi\rangle\nonumber \\
 & \quad & +\frac{1}{2}\left(\dot{\tilde{s}}+\tilde{s}^{2}\dot{s}\right)m\sum2q\frac{s^{(q)}}{q!}\langle\tilde{\phi}|e^{\tilde{S}}\left(\delta^{\dagger}\right)^{m+q-1}|\phi\rangle\bigr]\nonumber \\
 & = & i\sum_{m=2}-\frac{1}{m-1!}s^{(m)}\bigl[2\tilde{s}\dot{s}\langle\tilde{\phi}|e^{\tilde{S}}\left(\delta^{\dagger}\right)^{m}|\phi\rangle\nonumber \\
 & \quad & +\left(\dot{\tilde{s}}+\tilde{s}^{2}\dot{s}\right)(\Omega-m+1)\langle\tilde{\phi}|e^{\tilde{S}}\left(\delta^{\dagger}\right)^{m-1}|\phi\rangle\bigr]\nonumber \\
 & \quad & +i\left(\dot{\tilde{s}}+\tilde{s}^{2}\dot{s}\right)\sum_{m,q>1}\frac{s^{(m)}s^{(q)}}{(m-1)!(q-1)!}\langle\tilde{\phi}|e^{\tilde{S}}\left(\delta^{\dagger}\right)^{m+q-1}|\phi\rangle.\end{eqnarray}

\subsubsection{Result from RPA calculations}

\begin{figure}
\begin{centering}
\includegraphics[width=10cm]{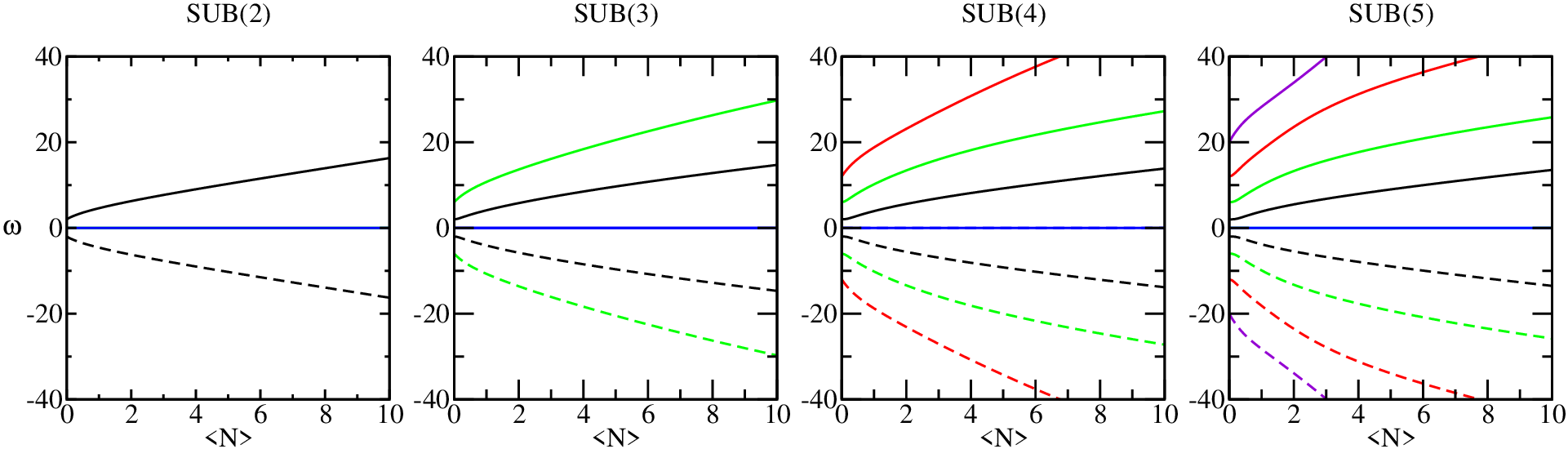}
\par\end{centering}

\caption{The RPA frequencies for the normal ECCM for $\Omega=10$ at various
levels of truncation\label{fig:ECCMRPA}}

\end{figure}

\begin{figure}
\begin{centering}
\includegraphics[width=10cm]{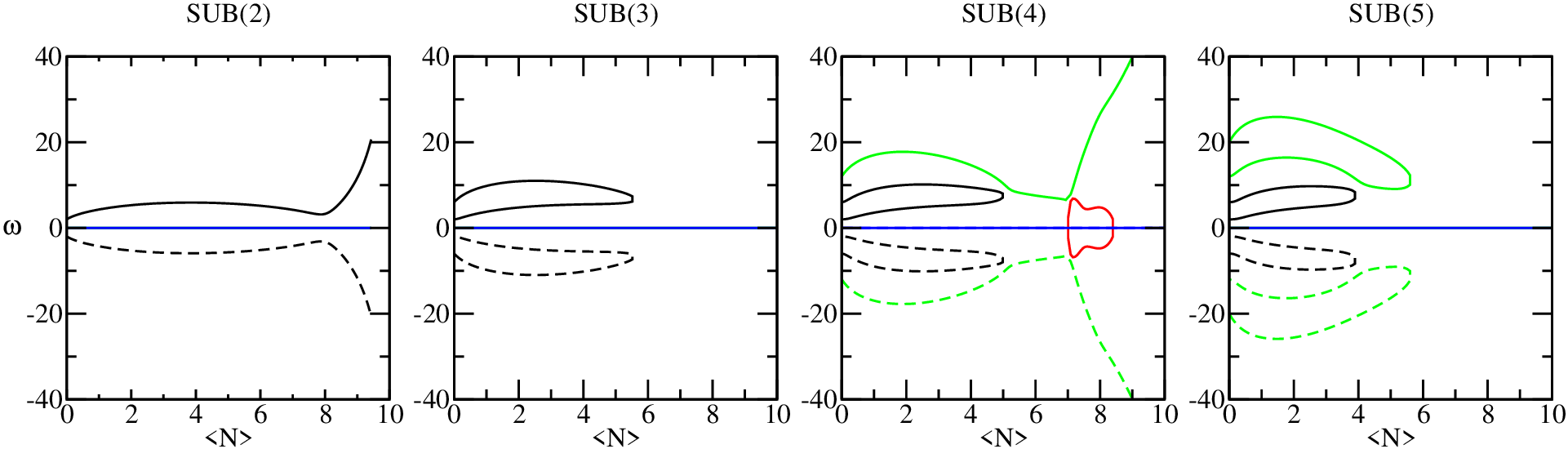}\caption{The RPA frequencies for the quasiparticle ECCM for $\Omega=10$ at
various levels of truncation\label{fig:qpECCMRPA}}

\par\end{centering}

\end{figure}

We have preformed RPA calculation for all of the approximation schemes
discussed above; we shall concentrate on two of most general schemes,
ECCM in normal operators and quasiparticle ECCM, Figs.~\ref{fig:ECCMRPA}
and \ref{fig:qpECCMRPA}, respectively. These two figures should be
looked at in conjunction with \ref{fig:Evsnc} and \ref{fig:Evsnb}.
In each of these we concentrate on those states with the same symmetry
as the ground state: our operators are only zero angular momentum
pairs.

From both these figures, we note that we do see the expected zero
modes for excitations breaking particle number, but remaining on the
solution branch being studied. These are of course doubly-degenerate.
As we increase the level of truncation, more-and-more other modes
come in. These also break particle number, as can be seen e.g. at
$\langle N\rangle=0$, where the only excited states available are
those with different particle number, e.g., the $N=2$ states for
SUB(2), $N=2+4$ for SUB(3), etc. The excitations energies are of
course also identical for the empty shell for the two schemes, since
they coincide at that point. Things change as we move away towards
the middle of the shell. Whereas the standard ECCM gives real frequencies,
the quasiparticle ECCM solutions meet in pairs (and turn complex),
showing a break-down of the approximation scheme. Also, the break-down
point seems to move to smaller $N$ as we increase the level of truncation.
It also sheds light on the rather different behaviour for even and
odd truncation levels seen before: for an even truncation scheme,
we always have one odd pair of modes, taking away the pair of zero
modes, whereas for an odd truncation level all non-zero modes can
and will collide in pairs.

\section{Discussion}

We have shown that one can formulate a number of CCM based methods
for application to problems with pairing. These have been analysed
for a rather simple model, and all seem to suffer from some problems.
The particle ECCM is the most stable method for low order calculations,
but all methods seem to suffer some drawback. The real surprise is
that the low order particle ECCM fails as badly as it does--we would
have expected it to work best. No detailed understanding of these
results exists, but they are clearly linked to the restoration of
particle number required in these calculations.

Clearly the exact integrability of the underlying model may also play
a role in these results, but one feature will persist: the restoration
of the broken particle number means that the exact solutions have
zero SUB(1) pairing, suggestive of the fact that different solutions
lie on different branches of CCM solutions. This will also lead to
instabilities at some order--we do expect that for real system with
pairing in large model spaces the problems are much less pronounced,
but the comparison of SUB2 NCCM and ECCM seems to be a promising aspect
of any approach, as is the study of the particle number fluctuations,
which seems to be the best tool to highlight unphysical solutions.

If we were to try and apply an improved many-body method to fermions,
e.g., confined in a harmonic trap, any one of the methods discussed
here is probably a good choice, as long as we don't push the order
of approximation too high; we must, however, carefully monitor the
behaviour of fluctuation in particle number and the local-harmonic
modes with ground state symmetry.

\section*{References}{}

\bibliography{CCM}

\end{document}